\documentclass[jgrga]{agu2001}
\usepackage{graphicx}
\titlerunninghead{DIURNAL POLAR MOTION DETECTION BY RING LASERS}
\authorrunninghead{SCHREIBER ET AL.}
\catcode`\@=11
\def\jheadline{\hbox to\textwidth{\iftitle%
\hfill\titlepageheadlinefont
\uppercase{JOURNAL OF GEOPHYSICAL RESEARCH, Vol.~109, B06405,}
doi:10.1029/2003JB002803, 2004\hfill%
\else\ifodd\c@page
{\hfill\headlinesize\headtextfont\theauthors:\ \ \thetitle}%
\hfill\llap{\foliofont B06405}%
\else\rlap{\foliofont B06405}\hfill%
{\headlinesize\headtextfont\theauthors:\ \ \thetitle}%
\hfill\fi\fi}} \def\cpr@year{2004}
\catcode`\@=12
\def\thecccline{%
0148-0227/04/2003JB002803/\$09.00}

\authoraddr{K.~U. Schreiber,
Technische Universit\"{a}t M\"{u}nchen,
Forschungs\-einrichtung Satellitengeod\"{a}sie
Fundamentalstation Wettzell,
D-93444 K\"{o}tzting, Germany.
(e-mail: schreiber@wettzell.ifag.de)}

\authoraddr{A. Velikoseltsev,
Technische Universit\"{a}t M\"{u}nchen,
Forschungs\-einrichtung Satellitengeod\"{a}sie
Fundamentalstation Wettzell,
D-93444 K\"{o}tzting, Germany.
(e-mail: alex@wettzell.ifag.de)}

\authoraddr{M. Rothacher,
Technische Universit\"{a}t M\"{u}nchen,
Forschungs\-einrichtung Satellitengeod\"{a}sie
Arcisstr. 21,
D-80333 M\"{u}nchen, Germany.
(e-mail: markus.rothacher@bv.tu-muenchen.de)}

\authoraddr{T. Kl\"{u}gel,
Bundesamt f\"{u}r Kartographie und Geod\"{a}sie,
Fundamentalstation Wettzell,
D-93444 K\"{o}tzting, Germany.
(e-mail: kluegel@wettzell.ifag.de)}

\authoraddr{G.~E. Stedman,
Department of Physics and Astronomy,
University of Canterbury,
Private Bag 4800, Christchurch, New Zealand.
(e-mail: geoffrey.stedman@canterbury.ac.nz)}

\authoraddr{D.~L. Wiltshire,
Department of Physics and Astronomy,
University of Canterbury,
Private Bag 4800, Christchurch, New Zealand.
(e-mail: david.wiltshire@canterbury.ac.nz)}

\begin{document}


\title{Direct measurement of diurnal polar motion\\ by ring laser gyroscopes}

\authors{K. U. Schreiber,
A. Velikoseltsev,
M. Rothacher}
\affil{Forschungs\-einrichtung Satellitengeod\"{a}sie,
Technische Universit\"{a}t M\"{u}nchen,
K\"{o}tzting, Germany.}
\authors{T. Kl\"{u}gel}
\affil{Bundesamt f\"{u}r Kartographie und Geod\"{a}sie,
Fundamentalstation Wettzell,
K\"{o}tzting, Germany.}
\authors{G. E. Stedman
and D. L. Wiltshire}
\affil{Department of Physics and Astronomy,
University of Canterbury, Christchurch, New Zealand.}


\begin{abstract}

We report the first direct measurements of the very small effect of forced
diurnal polar motion, successfully observed on three of our large ring lasers,
which now measure the instantaneous direction of Earth's
rotation axis to a precision of 1 part in $10^{8}$ when averaged over a
time interval of several hours. Ring laser gyroscopes provide a new
viable technique for directly and continuously measuring the position
of the instantaneous rotation axis of the Earth and the amplitudes of
the Oppolzer modes. In contrast, the space geodetic techniques
(VLBI, SLR, GPS, etc.) contain no information about the position of the
instantaneous axis of rotation of the Earth, but are sensitive to the
complete transformation matrix between the Earth-fixed and inertial
reference frame. Further improvements of gyroscopes
will provide a powerful new tool for studying the Earth's interior.
\end{abstract}


\begin{article}
\section{Diurnal Polar Motion}

The Earth's angular velocity vector varies in time, both in direction
and magnitude. Changes in magnitude correspond to a variation of the length
of day (LOD) with respect to atomic clocks, amount up to a few milliseconds
and consist mainly of various long period and seasonal components from solid
Earth tides and the interaction with atmosphere and ocean and small diurnal
and semi-diurnal terms from ocean tides.

The direction of the Earth's rotation axis also varies with respect to
both space- and Earth-fixed reference systems. The principal component
with respect to the Earth-fixed frame is the well-known Chandler wobble,
with an amplitude of 4--6m at the poles and a period of about 432 days.
This is a free mode of the Earth, i.e. it would
still be present in the absence of the external gravitational forces. It
is believed that the Chandler wobble would decay due to dissipative
effects in the Earth's interior, were it not continually excited by
seismic activity and by random noise of the atmosphere.

\begin{figure}
\noindent\includegraphics[width=20pc]{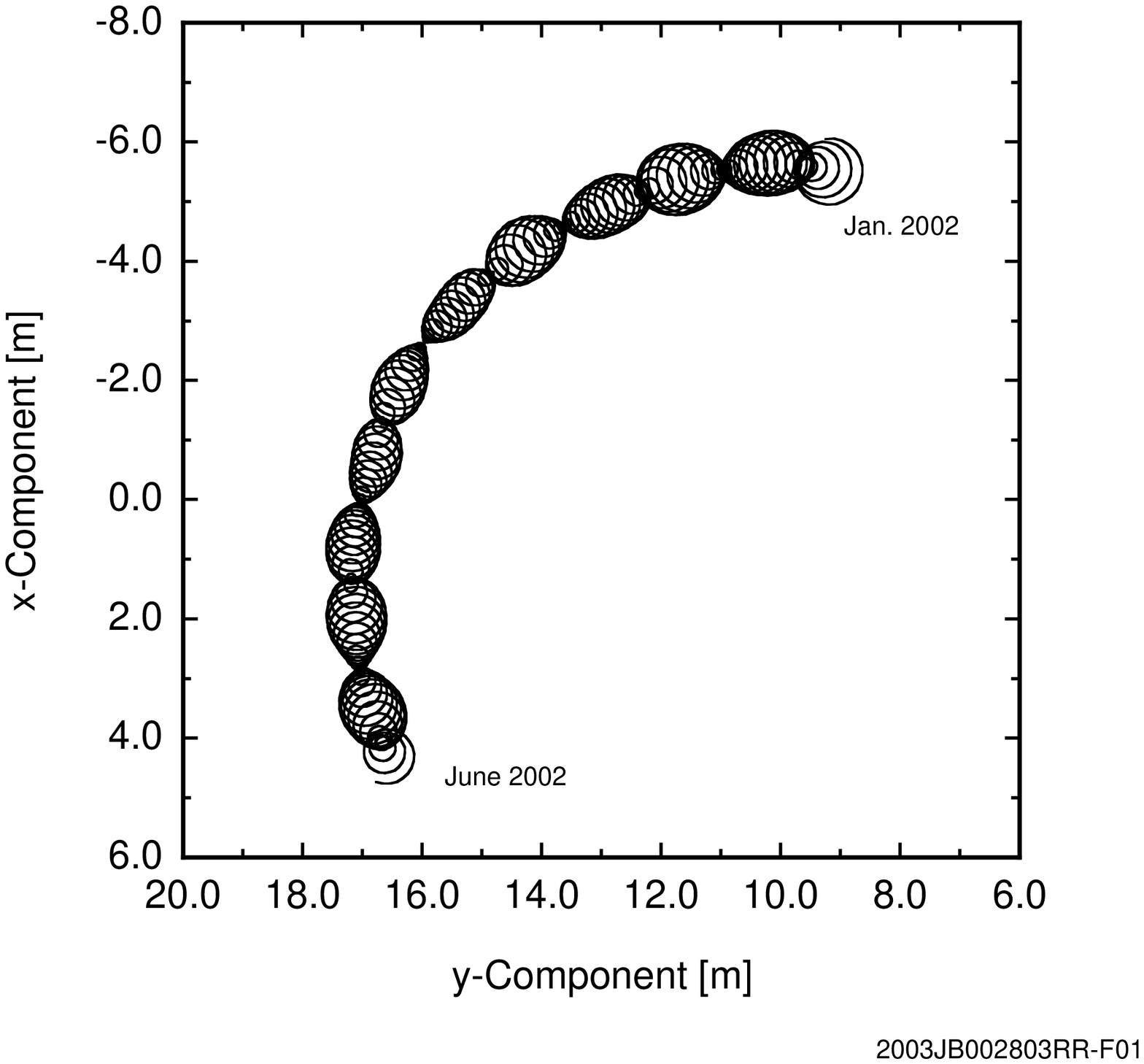}
\caption{{The behaviour of the Earth's instantaneous axis in an
Earth-fixed frame (w.r.t. the Conventional International Origin) in the period
of January to June 2002. The large circle is the Chandler
wobble. Superimposed are the smaller circles of the Oppolzer terms:
diurnal variations whose dominant amplitude varies over half a month.}}
\label{dpm}
\end{figure}

The Chandler wobble is overlaid by daily variations
whose amplitudes are an order of magnitude smaller,
some 40--60cm at the Earth's surface (c.f. figure~\ref{dpm})
\citep{mclure}, \citep{frede}. These Oppolzer terms arise from external
torques due to the gravitational attraction of the Moon and Sun. Since the
Earth is an oblate spheroid with an equatorial bulge which is inclined to
the plane of the ecliptic, the net gravitational torque of the Moon and Sun
on different parts of the Earth's surface does not exactly cancel out as
it would if the Earth were a perfect sphere.

In an Earth-fixed reference system, such as that of a ring laser fixed to
the Earth's surface, the forced retrograde diurnal polar motion is best viewed as a principal
mode -- the so-called ``tilt-over mode'' ($K_1$) -- with the period of
exactly one sidereal day ($23.93447$ hours), whose amplitude is modified as the
angles and distances between the Earth, Moon and Sun vary over the course of
their orbits. The complete spectrum of nutation modes can be understood as
the beat frequencies of the tilt-over mode with frequencies corresponding to
relevant orbital parameters: half a tropical month, half a tropical year, the
frequency of perigee etc. The beat periods are clustered around one sidereal
day. The $O_1$ and $P_1$ modes, with beat periods of $25.81934$ and $24.06589$
hours, have the largest amplitudes after the $K_1$ mode, and arise from the
change in angle between the Earth's equatorial bulge and the Moon and Sun
respectively. (See \citep{MM} for basic theoretical details.).

The spectrum of Oppolzer terms has the beauty that since it arises from
external gravitational torques, the frequencies are known precisely
but the amplitudes, which depend on the properties of the Earth such as
its inertia tensor, elasticity and liquid core
are not so precisely understood, and could potentially reveal much of
geophysical interest. To date models of the Earth's interior have been
built up which are broadly consistent with VLBI measurements of the
instantaneous orientation of the Earth's rotation axis. However, direct
routine measurements of the amplitude of the Oppolzer terms with ring
lasers would open up a new field of geophysical studies. In this paper
we report a first step in that direction.

\section{The Effect of Polar Motion on Ring Laser Gyroscopes}

Ring lasers measure absolute rotation~\citep{geoff}.
In the experimental set up two laser beams propagate in opposite directions
around a closed path. If the instrument rotates the effective path-length is
slightly shorter for the counter-rotating beam. In an active laser cavity,
as is the case for our instruments, lasing is achieved when an integral number
of wavelengths circumscribe the ring perimeter. Since the path length is
slightly different for the co-rotating and the counter-rotating beams
the lasing frequencies are also slightly different in each case and the
beat frequency of the two laser beams, the Sagnac frequency is readily
measurable. For an Earth-fixed rectangular cavity of perimeter length $P$,
and area $A$, with normal $\bf n$ the Sagnac frequency is given by
\begin{equation}
\delta f=\frac{4 A}{\lambda P}\; {\bf n \cdot \Omega } , \label{eq1}
\end{equation}
where $\bf\Omega$ is the instantaneous angular velocity vector
(referred to an inertial frame), and $\lambda$ is the wavelength
of the laser beams in the absence of rotation.
The Sagnac frequency will vary with:
(i) changes in the instantaneous position of the Earth's
rotation axis and its angular velocity, which alter $\bf\Omega$; (ii) tilts
from solid Earth tides and tidally
induced ocean loading changes, which alter the local normal, $\bf n$;
(iii) any changes to the scaling factor $A/P$, which could result
from local thermal or pressure induced variations in the dimensions of the
ring laser; and (iv) any changes in the refractive index seen by each
beam in the laser cavity, as might result from changes in gas composition.

Although different mechanisms contribute to changes in $\delta f$, their
individual effects can be separated. Firstly, unwanted effects from thermal
expansion and the like can be minimised by suitably isolating the ring
lasers and by building ring lasers with the largest possible ratio $A/P$
to increase their sensitivity. Likewise variation in refractive index is
an engineering design question and while problems such as tiny cavity leaks
may give slow drifts in $\delta f$ they would not have strongly periodic
signatures. Finally, tilts from solid Earth and ocean tides can be readily
distinguished from diurnal polar motion, by tiltmeter measurements on
the ring laser.

\callout{Table~1} summarizes the performance of some of the largest existing
ring lasers. The fourth column represents the experimentally obtained
instrumental resolution (averaging time of 3~hours), while the last column shows
the maximum amplitude of diurnal polar motion (DPM) as measured at each
ring laser site. While the longitude of a ring laser location determines the
phase of the polar motion signal, the latitude defines the projection of the
rotation vector onto the ring laser normal and therefore determines the amount of the
amplitude of the polar motion signal that is mapped into the ring laser
measurements by equation~\ref{eq1}.

\section{Details of a Diurnal Polar Motion Model}

McClure carried out a detailed investigation of diurnal polar motion,
using a purely elastic deformable Earth as the basic theoretical model \citep{mclure}.
The additional correction for liquid-core effects has been shown to produce
no significant differences for the motion of the rotation
axis \citep{wahr,brzezinski}. Adopting body--fixed
coordinates, the instantaneous position of the rotation axis is
determined as a linear superposition of sinusoidal Oppolzer modes.
By convention, the modes are expressed in terms of fundamental arguments
corresponding to particular orbital parameters. Specifically,
the longitude diurnal variation, $\Delta \varphi$, and the obliquity
diurnal variation, $\Delta \epsilon$, are given respectively by
\begin{equation}
\Delta \varphi = \sum_{i} -A_{i}\sin(\phi_{M}+\sum_{j}N_{ij}F_{j}),
\end{equation}
\begin{equation}
\Delta \epsilon = \sum_{i} A_{i}\cos(\phi_{M}+\sum_{j}N_{ij}F_{i})
\label{eq3}
\end{equation}
where $\phi_{M}$ is the Greenwich mean sidereal hour angle, $A_i$ are
amplitudes and $N_{ij}$ are integer coefficients (as given in
\citep{brzezinski}) which multiply the fundamental arguments like for example
\begin{eqnarray}
F_{1} \equiv l & = &134.^{\circ}96340251+171791592.''2178t \nonumber \\
 && +31.''8792t^{2}+0.''051635t^{3},
\end{eqnarray}
the mean anomaly of the Moon. Similar expressions are obtained for the
mean anomaly of the Sun, the difference between the mean longitude of
the Moon and the mean longitude of the ascending node of the lunar
orbit, the mean elongation of the Moon from the Sun and finally the mean
longitude of the ascending node of the Moon.
In all these expressions, the time $t$ is measured in Julian Centuries
of 36525 days of 86400 seconds since J2000. Furthermore, the Greenwich
mean sidereal hour angle $\phi_{M}$ relates to the Greenwich Mean
Sidereal Time (GMST) as follows:
\begin{eqnarray}
\phi_{M}&=&\hbox{GMST} \frac{2 \pi}{86400} + \pi + 2 \pi d_{u} ,\label{phim}\\
\hbox{GMST}&=& 24110^{s}.54841+8640184^{s}.812866T_{u} \nonumber \\
&& +0^{s}.093104T^{2}_{u}-6^{s}.2\cdot10^{-6}T^{3}_{u},
\end{eqnarray}
where $T_{u}=d_{u}/36525$, with $d_{u}$ the number of days elapsed since
12h UT1 January 1, 2000. Finally, the longitude $\lambda$ of each ring
laser must be added to the Greenwich mean sidereal angle (\ref{phim}) to
synchronise observations.
\begin{table}[t]
\caption{\label{tab1} The sensitivity of some of our ring lasers
and the theoretical maximum amplitude of the diurnal polar motion effect at these
instruments.}
\begin{flushleft}
\begin{tabular}{|c|c|c|c|c|}
\tableline
Ring laser &Area & $f_{Sagnac}$ & Gyroscope Sensitivity & $DPM_{max}$\\
&$\mbox{m}^{2}$ & Hz & $\delta \Omega/\Omega$ & $\delta \Omega/\Omega$\\
\tableline
C-II&1 & 79.4 & $1 \cdot 10^{-7}$  & $1 \cdot10^{-7}$\\
G&16 & 348.6  & $1 \cdot 10^{-8}$  & $9.8 \cdot10^{-8}$\\
UG1&367 & 1512.8 & $3 \cdot 10^{-8}$  &  $1 \cdot10^{-7}$\\
\tableline
\end{tabular}
\end{flushleft}
\end{table}


The tables given in~\citep{mclure},\citep{brzezinski} provide up to 160
values of diurnal polar motion amplitudes and coefficients.
We will keep only the largest 6 coefficients, as all the remaining
coefficients together contribute less than
$1$ milliarcsecond (mas) to the amplitude of polar motion and are
not yet within the resolution obtained with a gyroscope.
Calculations with these assumptions superimposed on the
official International Earth Rotation Service (IERS)
C04 polar motion series~\citep{iers} were used to compute figure~\ref{dpm}.
This C04 series refers to the Celestial Intermediate Pole (CIP)~\citep{cip}
defined to be free of diurnal polar motion terms, as opposed to
the instantaneous axis of rotation.
One can clearly see the diurnal circles of the rotation axis as it
progresses through a larger cycle of the Chandler wobble. The resulting
``latitude oscillations'' have two main components with a periodicity of
nearly 24 hours and approximately 14 days. The maximum amplitude
represents 0.02 seconds of arc at the pole. We wish to point out that besides
the here discussed forced diurnal polar motion there are many more effects that have subdaily or daily
contributions to Earth rotation (e.g. from atmosphere, ocean). However at the current level of performance
of our ring laser gyroscope they are not yet observable. This also
applies for length of day variations.

\section{Direct Observation of Polar Motion by Large Ring Lasers}

The effect of Earth tides and ocean loading as small periodic variations has
already been successfully identified in the time series of the Sagnac
frequency of the small C-II ring laser~\citep{TiltTide} located in an
underground cavern in Christchurch (New Zealand). An unambiguous detection of
polar motion has now been provided by our other ring lasers, G and UG1, the
results from G being the most accurate and definitive. G is a semi-monolithic
square HeNe ring laser of 16~$\mbox{m}^{2}$ area constructed from Zerodur and
located in a thermally stable underground laboratory at the Fundamentalstation
Wettzell (Germany)~\citep{g-paper}.
The G ring laser was operated for most of the time in the year 2002. The
most stable conditions ever obtained from any of our large ring lasers
were met between July 24 and October 27, corresponding to the days
205 and 300 of the year 2002.
The overall drift of the raw data over the entire time is
as low as 2 mHz or expressed as an relative error $\Delta f/f < 6 \cdot
10^{-6}$ over 95 days, to our knowledge a world record for gyroscope
stability.

\begin{figure}
\noindent\includegraphics[width=20pc]{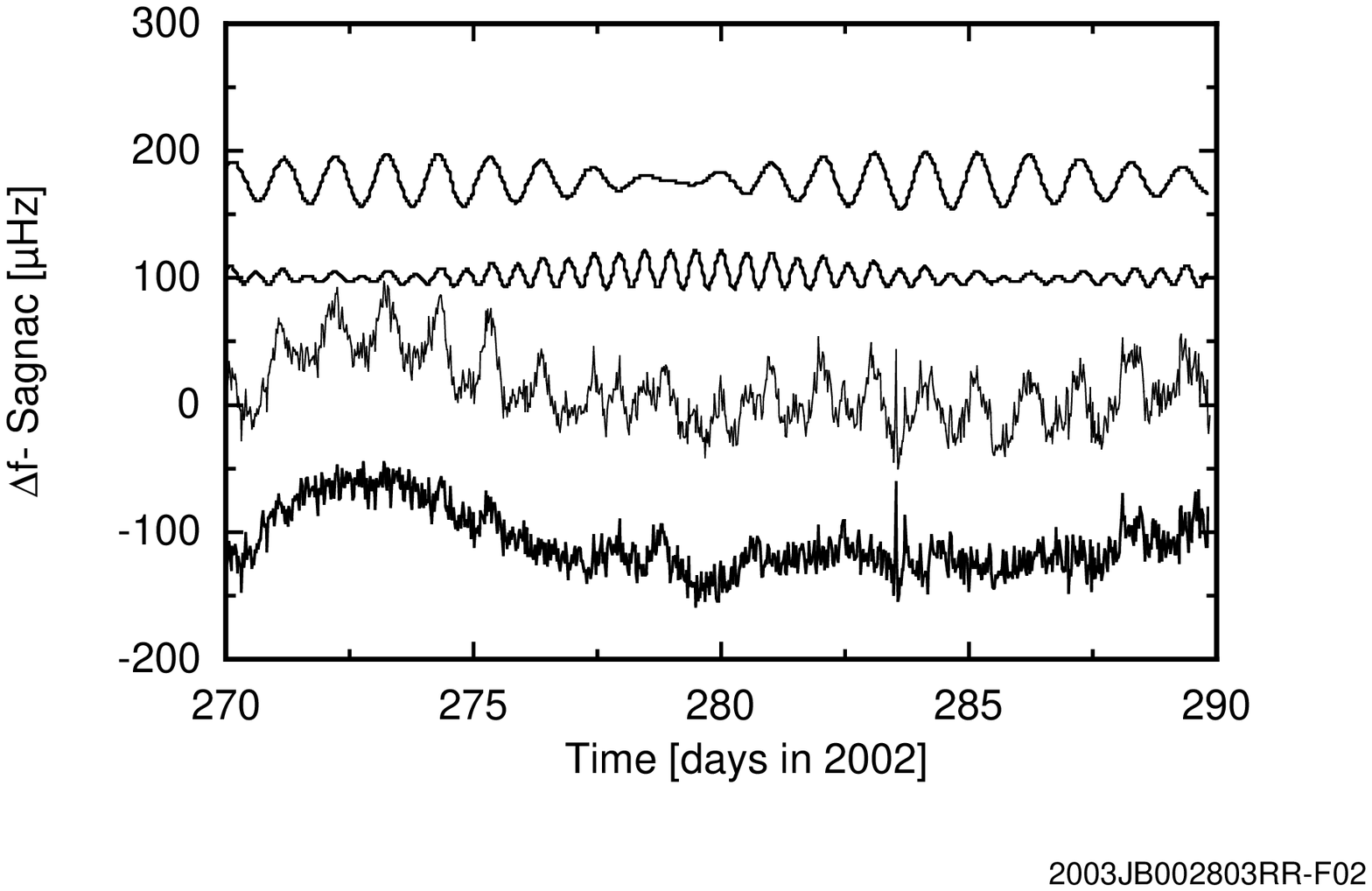}
\caption{Time series of a section of the Sagnac frequency
taken by the G ring laser. The computed contributions of the Earth body
tides and diurnal polar motion to the Sagnac frequency are shown offset by
100 $\mu$Hz and 180 $\mu$Hz respectively at the top. The resultant time
series with both corrections applied is shown offset at the bottom.}
\label{geo}
\end{figure}

A section of the observed time series -- data taken between
days 270 and 290 -- is shown in \callout{figure~\ref{geo}}. A constant value of
348.635 Hz has been subtracted from each measurement. In addition to the
raw ring laser data one can find the corresponding tilt signatures from
the body tides~\citep{Agnew} and the diurnal polar
motion~\citep{mclure,brzezinski}, both converted to the respective
variations of the Sagnac frequency according to equation~\ref{eq1} and
offset by 100 and 180~$\mu$Hz
for better illustration in the diagram. The residual data set is displayed
offset by -100~$\mu$Hz after the corrections were applied and shows a
good agreement between the measurements and the models adopted. What remains is
the signature from instrumental drift and some transient pertubations
of still unknown geophysical origin.

\begin{figure}
\noindent\includegraphics[width=20pc]{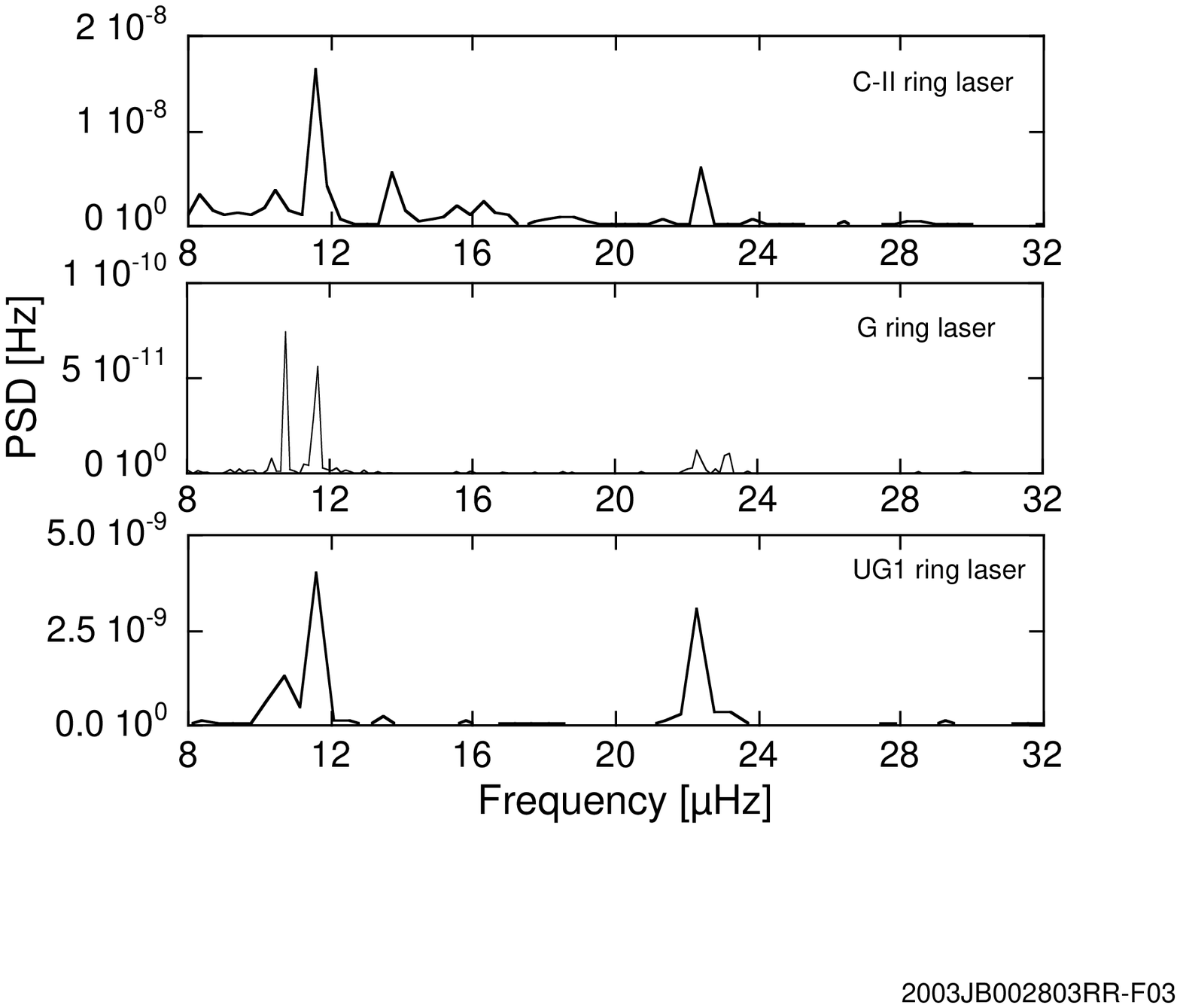}
\caption{Power spectra of ring laser time series from C-II, G and
UG1. The expected contributions from solid Earth tides and contributions
to polar motion from the $K_{1}$ and $O_{1}$ mode are clearly present in all
measurements. The spectrum of G reveals the most details because the
data set is much longer and the instrument is mostly free from
backscatter and drift.}
\label{C2-G-UG1}
\end{figure}

We have computed power spectra of ring laser measurements from various
available time series of different lengths of 3 different ring lasers,
C-II, G and UG1. \callout{Figure~\ref{C2-G-UG1}} shows the result.
There are 2 groups of signals evident in the spectrum of the G ring, one
at a period around 12 hours 25 minutes and one at a period of around one
day. In the latter group we can identify the $K_{1}$ and $O_{1}$ mode
contributions to diurnal polar motion. Their periods of
23.9345 and 25.8195 hours \citep{frede} correspond to frequencies of
11.606 and 10.758~$\mu$Hz. Our measurements yield frequencies of 11.665
and 10.754~$\mu$Hz, which agree well with theory within the spectral
measurement resolution of 129 nHz. The $P_{1}$ mode also
has an expected amplitude well within current resolution, but since its
period is very close to that of $K_1$ longer time series are
required to separate out its peak in the spectrum.

The group of higher frequencies is due to variations of the orientation
of the ring laser on the Earth surface induced by solid Earth
tides. Again the frequencies of 23.204 and 22.293~$\mu$Hz are within the
resolution of the expected $S_{2}$ and $M_{2}$ tides signals.
The Earth tides contribute little to the diurnal terms, because north
directed diurnal tidal tilts are close to zero in mid-latitudes.

The best available time series of UG1 and C-II are much shorter.
Therefore the resolution of their spectra is not so high. In addition,
C-II shows a significant masking of the diurnal polar motion signal,
probably due to backscatter induced frequency pulling as an instrumental
artefact~\citep{stabil}. Furthermore, the power spectral densities obtained for
daily polar motion and solid Earth tides as measured with the C-II
ring laser are about one order of magnitude larger than expected.
By contrast, for UG1 and G this is not the case. We believe that this
is related to the backscatter induced pulling of the Sagnac frequency
which is strong in C-II but negligible for UG1 and G.
Due to long timescale instrumental drift frequencies below
5~$\mu$Hz (G) or 8~$\mu$Hz (UG1) were fully cut off by the application
of an 8th degree Butterworth high pass-filter. However, one can clearly
identify diurnal polar motion signals and solid Earth tides in all
three instruments. The tidal signal is stronger at Christchurch due to
local ocean loading on Banks Peninsula.

\begin{figure}
\noindent\includegraphics[width=20pc]{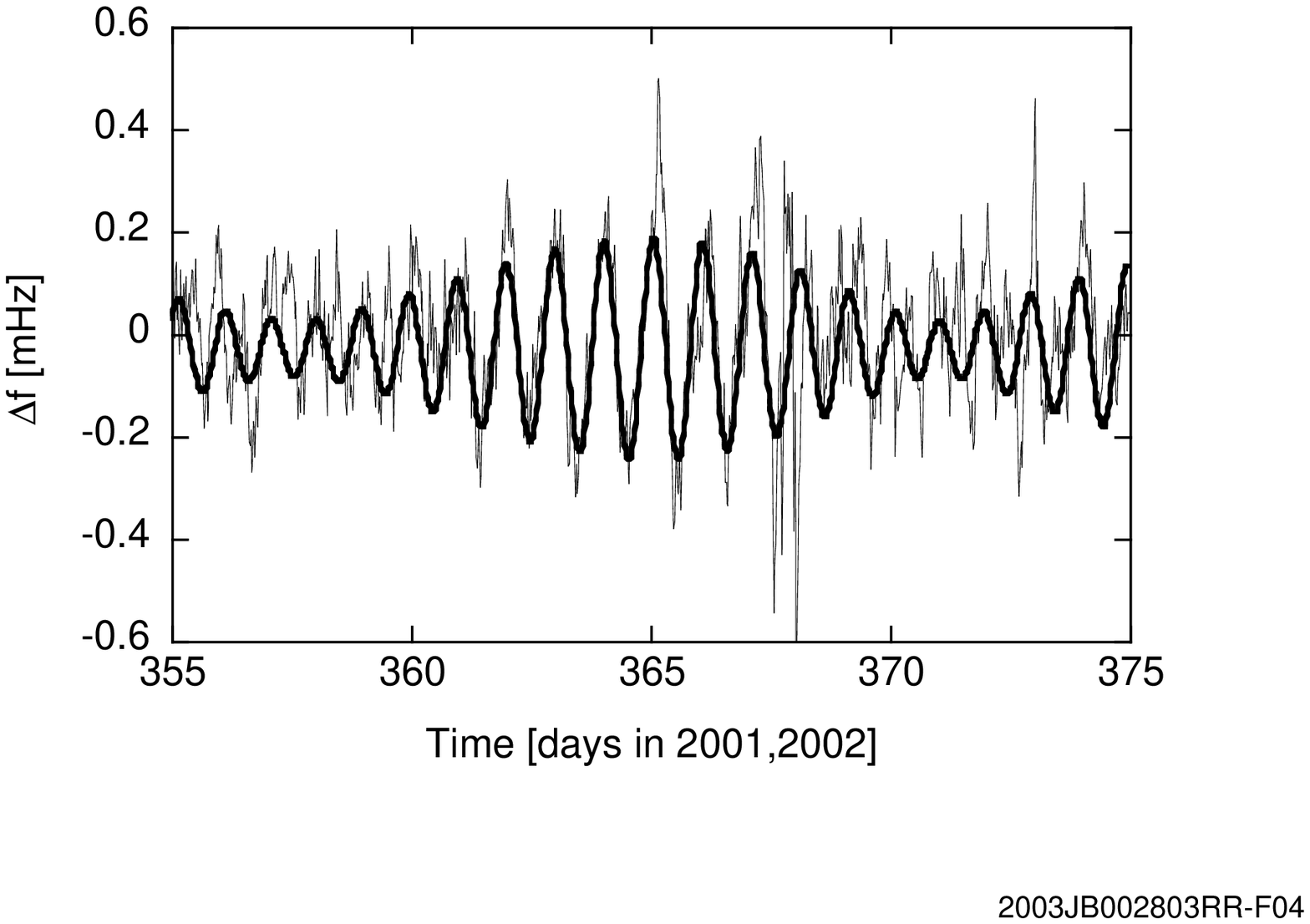}
\caption{An example of a time series of a measurement of diurnal
polar motion carried out on the UG1 ring laser. Because of slow instrumental
drift the data were band-pass filtered. The computed theoretical polar
motion signal was treated similarly and is superimposed on the
Sagnac frequency measurements. The data set was also corrected for Earth
tides and ocean loading.}
\label{ug1-time}
\end{figure}

The larger a ring laser in a given location is, the higher is its
sensitivity to such small signals. \callout{Figure~\ref{ug1-time}} shows a time
series from the 367~$m^{2}$ UG1 which we believe is the largest working
ring laser gyroscope in the world~\citep{ug1}.
It dramatically shows one of the advantages of ring lasers, namely the
very high time resolution of the measurements, making this technique very
promising for the study of subdaily variations in Earth rotation.
In addition continuous
observations are possible over long time intervals. In contrast gyroscopes
are local sensors and the data may contain contributions from local effects
which are not yet understood. Indications for this are apparent in
\callout{figure~\ref{ug1-time}} as well.

Since the raw data of the UG1 ring laser shows far more drift than the G
ring laser, the data were band-pass filtered with cut-off frequencies at 5
and 40~$\mu$Hz. The computed polar motion signal was treated in the same
way in order to avoid artefacts from phase shifts caused by the filter
process, and is superimposed on the measurement data in
\callout{figure~\ref{ug1-time}}. The data set has been corrected for Earth
tides and ocean loading~\citep{Agnew}.
Local disruptions of the ring laser setup have not been reduced from the
data set. This plot shows the impressive sensitivity of modern
large ring laser gyroscopes applied to the field of geophysics.

\section{Summary}

The sensitivity and performance of large ring lasers has improved
significantly in recent years with the construction of the C-II, G0, G
and UG1 gyroscopes. Today it is possible to measure variations in the
location of the rotational pole of the Earth to within a few
centimetres. As our instruments progressively improve in
stability we expect evidence of further geophysical signals with
periods of well over a day in the future. Since both our ring laser
laboratories in Germany and New Zealand are nearly at antipodal points
we unfortunately cannot yet distinguish the different effects of the
components $\Delta\varphi$ and $\Delta\epsilon$ of the polar motion.
This will require at least one additional high quality ring laser site
displaced 90 degrees in longitude from our current laboratories.

It is important to note that over the past 30 years the theoretical
model of forced diurnal polar motion has been developed and is used to
reduce these contributions from VLBI measurements. There are still some
uncertainties remaining, since the theoretical models use some
simplifications to account for a deformable Earth. Large ring
laser gyroscopes provide a new and independent technology which can measure the
amplitudes and frequencies of the forced modes of Earth rotation directly
and monitor polar motion to unprecedented high temporal resolution.
This effectively provides a direct probe of the Earth's moment of inertia.
Therefore we expect that future improvements in ring laser technology
will lead to quantitative improvements in the nutation models themselves,
and provide a new tool for studying aspects of the Earth's interior.

\begin{acknowledgments}

These combined ring laser results were possible because of a collaboration of
Forschungs\-einrichtung Satellitengeod\"{a}sie, Technische
Uni\-ver\-si\-t\"{a}t M\"{u}n\-chen, Germany, University of Canterbury,
Christchurch, New Zealand and Bundesamt f\"{u}r Kartographie and Geod\"{a}sie,
Frankfurt, Germany. University of Canterbury research grants, contracts of
the Marsden Fund of the Royal Society of New Zealand and also grants from
the Deutsche Forschungs\-gemeinschaft (DFG) are gratefully acknowledged.
\end{acknowledgments}

\end{article}

\begin{thebibliography}{}
\bibitem[{\it Agnew}(1997)]{Agnew}
Agnew, D.~C. (1997),
NLOADF: A program for computing ocean-tide loading,
{\it J. Geophys.\ Res}.\ {\bf102}, 5109--5110.

\bibitem[{\it Brzezi\'nski}(1986)]{brzezinski}
Brzezi\'nski, A. (1986),
Contribution to the theory of polar motion for an elastic earth
with liquid core,
{\it Manuscripta Geodaetica} {\bf11}, 226--241.

\bibitem[{\it Capitaine et al.}(2002)]{cip}
Capitaine, N., et al.~(Eds.) (2002), {\it Proceedings of the IERS Workshop
on the Implementation of the New IAU Resolutions, IERS Tech.\ Note 29},
Int.\ Earth Rotation Serv., Cent.\ Bur., Frankfurt-am-Main, Germany.

\bibitem[{\it Dick and Richter}(2002)]{iers}
Dick, W.~R. and B. Richter~(Eds.) (2002),
IERS Annual Report 2001, 123 pp, Int.\ Earth Rotation Serv.,
Cent.\ Bur., Frankfurt-am-Main, Germany.

\bibitem[{\it Dunn et al.}(2002)]{ug1}
Dunn, R. W., D.~E. Shabalin, R.~J. Thirkettle, G.~J. MacDonald,
G.~E. Stedman, and K.~U. Schreiber (2002),
Design and initial operation of a 367 m$^2$\ rectangular ring laser,
{\it Appl.\ Opt}.\ {\bf41}(9), 1685--1688.

\bibitem[{\it Frede et al.}(1999)]{frede}
Frede, V. and V. Dehant (1999), Analytical versus semi-analytical
determinations of the Oppolzer terms for a non-rigid Earth,
{\it J. Geodesy} {\bf73}, 94--104.

\bibitem[{\it McClure}(1973)]{mclure}
McClure, P. (1973), Diurnal polar motion,
{\it GSFC Rep. X-529-73-259}, Goddard Space Flight Center, Greenbelt, Md.

\bibitem[{\it Moritz et al.}(1987)]{MM}
Moritz, H., and I.~I. Mueller (1987),
{\it Earth Rotation: Theory and Observation},
(Ungar, New York).

\bibitem[{\it Schreiber et al.}(1998)]{stabil}
Schreiber, K.~U., C.~H. Rowe, D.~N. Wright, S.~J. Cooper, and G.~E. Stedman
(1998), Precision stabilization of the optical frequency in a large ring
laser gyroscope,
{\it Appl.\ Opt}.\ {\bf37}(36), 8371--8381.

\bibitem[{\it Schreiber et al.}(2001)]{g-paper}
Schreiber, K.~U., A. Velikoseltsev, T. Kl\"ugel,
G.~E. Stedman, and W. Schl\"uter (2001),
Advances in the Stabilisation of Large Ring Laser Gyroscopes,
paper presented at 
the Symposium Gyro Technology, Univ. of Stuttgart,
Stuttgart, Germany.

\bibitem[{\it Schreiber et al.}(2003)]{TiltTide}
Schreiber, K.~U., G.~E. Stedman and T. Kl\"ugel (2003),
Earth tide and tilt detection by a ring laser gyroscope,
{\it J.\ Geophys.\ Res.}\ {\bf108}(B2), 2132, doi:10.1029/2001JB000569.

\bibitem[{\it Stedman}(1997)]{geoff}
Stedman, G.~E. (1997),
Ring-laser tests of fundamental physics and geophysics,
{\it Rep.\ Prog.\ Phys}.\ {\bf60}, 615--688.

\bibitem[{\it Wahr}(1981)]{wahr}
Wahr, J.~M. (1981),
The forced nutations of an elliptical, rotating, elastic and oceanless
earth,
{\it Geophys.\ J. R. Astr.\ Soc}.\ {\bf64}, 705--727.
\end{thebibliography}
\end{document}